\documentclass[twocolumn,showpacs,preprintnumbers,amsmath,amssymb]{revtex4}

\usepackage{graphicx}
\usepackage{dcolumn}
\usepackage{bm}
\usepackage{epsfig}
\usepackage[mathscr]{euscript}
\usepackage{color}

\newenvironment{Notes}
{\begin{quote}\small\tt Note to Ioan: \ }
{\end{quote}}

\newcommand{\beq}{\begin{equation}}
\newcommand{\eeq}{\end{equation}}
\newcommand{\bno}{\begin{Notes}}
\newcommand{\eno}{\end{Notes}\noindent}

\begin{document}

\title{Advancements in Milestoning II: Calculating Autocorrelation from Milestoning Data Using Stochastic Path Integrals in Milestone Space}

\author{Gianmarc Grazioli}
\author{Ioan Andricioaei}
 \email{andricio@uci.edu}
\affiliation{Department of Chemistry, University of California, Irvine, CA 92697}

\date{\today}

\begin{abstract}
\noindent
The Milestoning method has achieved great success in the calculation of equilibrium kinetic properties such as rate constants from molecular dynamics simulations. The goal of this work is to advance Milestoning into the realm of non-equilibrium statistical mechanics, in particular, the calculation of time correlation functions. In order to accomplish this, we introduce a novel methodology for obtaining flux through a given milestone configuration as a function of both time and initial configuration, and build upon it with a novel formalism describing autocorrelation for Brownian motion in a discrete configuration space. The method is then applied to three different test systems: a harmonic oscillator, which we solve analytically, a two well potential, which is solved numerically, and an atomistic molecular dynamics simulation of alanine dipeptide.      
\end{abstract}

\pacs{Valid PACS appear here}

\maketitle

\section{Introduction}
The calculation of time correlation functions from time series measurements made along molecular dynamics trajectories plays the same central role in kinetics as calculating partition functions from sets of molecular configurations and their respective energies in the realm of thermodynamics. To put the magnitude of this task into perspective, consider a simple system where 100 different configurations are possible, and a transition between any pair of these configurations is possible. In this simple system, there are over $1.7 \times 10^{13}$ different 10 step trajectories possible (100 choose 10) without even considering the fact that the same series of 10 configurations can occur with different transition times which makes the number of possible trajectories proliferate even further!  All important experimental properties can be calculated from time correlation functions measured from molecular dynamics simulations, but these effects are typically only measurable on timescales which are out of reach for brute force molecular dynamics. An example would be calculating RDCs (Residual Dipole Couplings) from NMR experiments from bond vector time correlation functions. The challenge of and demand for calculating kinetic properties from molecular dynamics simulations have caused it to become a major growth area in chemical physics \cite{dellago1998transition} \cite{dellago1999calculation}, leading to the development of several methods, spanning from early treatments using transition state theory (TST) \cite{Eyring} \cite{Wigner1938}, to more recently, transition path sampling (TPS) \cite{Bolhuis2002}, transition path theory (TPT) \cite{VandenEijnden2009}, and transition interface sampling (TiS) \cite{TiS}. A common strategy in measuring kinetics in molecular dynamics simulations is the measurement of fluxes of trajectories through hyperplanes in phase space or configuration space \cite{Elber2004} \cite{vanden2005} . More recently, the use of the hyperplanes in the Milestoning method has been generalized to subdividing phase space into Voronoi cells, where the milestones exist as the interfaces between cells \cite{voronoi}. Thus far, Milestoning has been used to calculate many useful properties, such as equilibrium flux values through the set of milestones, rate constants \cite{west2007}, and other equilibrium properties such as mean first passage times between states \cite{exactMstone}, but the method has never before been used to calculate non-equilibrium dynamical objects such as time correlation functions. In our first paper, Advancements in Milestoning I, we introduced a methodology for rapid calculation of transition time density functions between milestone hyperplanes, the central objects of milestoning calculations, by artificially pushing the system toward the target milestone and then re-weighting the distribution to recover the true transition time distribution \cite{grazioli1}. In this paper, we venture into this realm by introducing a method for calculating time correlation functions from milestoning data. In order to calculate autocorrelation from milestoning data, not only must we know the equilibrium flux values through each interface, we must also know the flux through each interface as a function of time and initial configuration. For this reason, it was necessary that we also introduce our stochastic path integral approach to calculating the time-dependent fluxes, in addition to the methodology for calculating time correlation functions from these time-dependent fluxes. 

\section{Theory}
\noindent 

\subsection*{Milestoning Theory}


A more in-depth overview of milestoning theory can be found in our first paper \cite{grazioli1}, or in \cite{west2007}, but let us review a few of the key premises upon which our method for calculating time correlation functions hinge. The quantity of most fundamental importance in milestoning is the flux through a given milestone, for which the equation is \cite{Elber2004}:

\begin{equation}
\label{m}
P_s(t) = \int_0^t Q_s(t')\left[ 1-\int_0^{t-t'}K_s(\tau)d\tau \right]dt' , \nonumber
\end{equation}

\begin{equation}
Q_s(t) = 2 \delta(t)P_s(0) + \int_0^t Q_{s\pm1}(t'')K^{\mp}_{s\pm1}(t-t'')dt''
\end{equation}
\noindent
where $P_{s}(t)$ is the probability of being at milestone $s$ at time $t$, (or, more specifically, arriving at time $t'$ and not leaving before time $t$ \cite{Elber2004}), and $Q_{s}(t)$ is the probability of a transition to milestone $s$ at time t.  $K_s(\tau)$ indicates the probability of transitioning out of milestone $s$ given an incubation time of $\tau$, thus $\int_0^{t-t'}K_s(\tau)d\tau$ is the probability of an exit from milestone $s$ anytime between $0$ and $t-t'$, which makes $1-\int_0^{t-t'}K_s(\tau)d\tau$ the probability of there \emph{not} being an exit from milestone $s$ over that same time period. Since the probability of two independent events happening concurrently is  the product of the two events, the equation for $P_s(t)$ is simply integrating the concurrent probabilities of arriving at milestone $s$ and not leaving over the time frame from time $0$ to $t$. Turning our attention towards the meaning of the first term, $Q_{s}(t)$, $2 \delta(t)P_s(0)$, simply represents the probability that the system is already occupying milestone $s$ at time $t = 0$, where the factor of 2 is present since the $\delta$-function is centered at zero, meaning only half of its area would be counted without this factor. $Q_{s\pm1}(t'')$ is the probability that the system transitioned into one of the two milestones adjacent to $s$ at an earlier time $t''$. $K^{\mp}_{s\pm1}(t-t'')$ is the probability of a transition from milestones $s\pm 1$ into milestone $s$. Thus the second term of the second line of equation 14 is another concurrent probability: the probability of the system entering an adjacent milestone at an earlier time, and then transitioning into milestone $s$ between time $t$ and $0$. It is important to note that all functions $P_s(t)$ and $Q_s(t)$ are calculated using the respective values of $K_s(\tau)$ between adjacent milestones, thus the set of $K_s(\tau)$ between all milestones of interest contains all the information needed to calculate kinetics using the milestoning method. It is also important to note that a $K$ function between two milestones $x = A$ and $x = B$, $K_{AB}(\tau)$, is simply a probability distribution representing the lifetime for the system remaining in state $A$ before transitioning to state $B$.

\subsection*{Time Correlation from Milestoning Data}

This approach aims to glean the time correlation function $C(t)$ of an observable from Milestoning data. The key insight into this method is the approximation of the continuous configuration space, which we define as $x$, as a discrete space of milestone configurations. Although the formalism presented below requires that the equilibrium distribution of configurations occupied, $f(x)$, is known, any successful Milestoning simulation yields the equilibrium flux through the set of milestones, and so this set of fluxes will serve as the equilibrium distribution of configurations in our discrete space. For the sake of clarity of notation, we will be limiting our derivation to observables which are a function of configuration $x$, but it should be noted that all developments presented herein can be easily generalized to observables which are a function of both position and velocity by considering our variable $x$ as a phase space coordinate. We begin with the usual definition for a time correlation function for time-ordered measurements of an observable that is a function of configuration, $A(x;t)$, arising from the equilibrium distribution of configurations, $f(x)$:

\begin{equation}
C(t) = \langle A(x,0)A(x,t) \rangle = \int A(x_0, 0)A(x, t) f(x) dx
\label{coft}
\end{equation}    
\noindent
where time t is the lag time between measurements. For time $t = 0$ the time correlation function has the lower limit $C(0) = \int A(x_0,0)A(x_0,0) f(x) dx = \langle A^2 \rangle$, the variance. On the opposite extreme, given an infinite relaxation time, the mean value of $x$ at time $t$ will be equivalent to the mean at equilibrium,  $\lim_{t \to \infty} \langle A(x,t) \rangle = \int A(x) f(x) dx$, which implies: $\lim_{t \to \infty} C(t) = \int A(x) \left( \int A(x) f(x) dx \right) f(x) dx = \int A(x) f(x) dx \int A(x) f(x) dx = \langle A \rangle^2$

So far, we have only discussed equilibrium probability distributions in configuration space, which we defined as $f(x)$, but let us now consider a time-dependent probability density function of configuration, which is a function of initial configuration $x(0)$. Keep in mind that time-dependent probability density functions such as these are the solutions to Fokker-Planck equations. Let us define this probability density function as $g(x, t ; x_0, 0)$, and express its mean value as a function of time and initial configuration, $\langle x(t, x_0) \rangle$, in the following manner:

\begin{equation}
\langle x(t, x_0) \rangle = \int x g(x, t ; x_0, 0) dx
\label{xg}
\end{equation}    
\noindent
Following suit, the expectation value of our observable $A$ as a function of time can be written as:

\begin{equation}
\langle A(x, t ; x_0, 0) \rangle = \int A(x) g(x, t ; x_0, 0) dx
\label{xg}
\end{equation}    
\noindent
We can now substitute $\langle A(x, t ; x_0, 0) \rangle$ for $A(x ,t)$ in the definition of a time correlation function:

\begin{equation}
C(t) = \int A(x) \left( \int A(x) g(x, t ; x_0, 0) dx \right) f(x) dx
\label{int1}
\end{equation}    
\noindent
As stated earlier in this section, our aim is to coarse grain the continuous configuration space of $x$ into a discrete space of milestone configurations, from which we can calculate a time correlation function. Our first step in constructing this model will be to approximate the outermost integral in $x$ with a sum over a discrete set of configurations $\{x_i\}$ multiplied by the equilibrium probability of finding the system in the configuration $i$. If we define the probability of the system being in configuration $x_i$ at time $t$ given an initial configuration $x_0$ as $P_i(t ; x_0)$, then given that our system will reach equilibrium at infinite time regardless of initial configuration, the equilibrium probability can be expressed as $P_i(\infty)$. Thus we arrive at our first discrete approximation of time correlation:

\begin{equation}
C(t) \approx \sum_i A(x_i) P_i(\infty) \left( \int A(x) g(x,x(0), t) dx \right) 
\label{intr}
\end{equation}     
\noindent
Our next task is to approximate the remaining integral in the equation with a sum over milestone states. Equation \ref{xg} gives us an expression for the mean value of $A(x)$ in a continuous space, given an amount of time elapsed $t$ and an initial configuration $x_0$. Now consider the case where $x$ can only occupy discrete values from the set $\left\{ x_s \right\}$. In this case, the integral in equation \ref{xg} is replaced by a sum in a weighted average expression where each discrete value of $x_i$ multiplied by its statistical weight as a function of time:

\begin{equation}
 \int A(x) g(x,x(0), t) dx \approx \sum_s A(x_s) P_s(t | x_0)
\end{equation}     
\noindent
Next, we substitute this weighted sum approximation into equation \ref{intr}: 

\begin{equation}
C(t) = \sum_i \left( A(x_i) P_i(\infty) \sum_s A(x_s) P_s(t | x_0) \right) 
\label{geq}
\end{equation}     
\noindent

Note that we have now arrived at a complete expression for a discrete approximation of time correlation, with the assumption that $P_s(t | x_0)$ and $P_i(\infty)$ can be obtained from milestoning calculations. Since the set of equilibrium fluxes, $P_i(\infty)$, have been calculated from milestoning simulations since the beginning, and we will introduce a novel method for calculating $P_s(t | x_0)$ from milestoning simulations in the Random Walk / Path Integral Methodology subsection later in the article, we are able to demonstrate that time correlation can indeed be calculated from Milestoning simulations.

\section{Analytical Solution for 1D Harmonic Oscillator}

In this section, we demonstrate the effectiveness of equation \ref{geq} in approximating the time correlation function for diffusion in a harmonic potential, for which there is an analytical solution. Our potential is defined as $V(x) = \frac{1}{2} k x^2$, and it's equilibrium distribution in $x$ is the Boltzmann distribution, $f(x) = e^{-\beta V(x)}$. The closed form expression for the time-dependent probability distribution for diffusion in a harmonic well is \cite{klausNotes}:  

\begin{equation}
p(x, t | x_0, 0) = \frac{1}{ \sqrt{2 \pi k_BT S(t)/k}} \exp \left[ -\frac{\left( x - x_0e^{-2t/\bar{\tau}} \right)^2}{2 k_BT S(t)/k} \right] 
\label{anap} 
\end{equation}
\noindent
where $S(t) = 1-e^{-4t/\bar{\tau}}$ and $\bar{\tau} = 2k_BT / kD$. 

Given this analytical expression for $p(x, t, | x_0, 0)$, we can obtain an analytical expression for $C(t)$ by substituting $p(x, t, | x_0, 0)$ into equation \ref{int1} for $g(x, x_i(0), t)$ and integrating. This yields the exact time correlation function $C(t)$ for diffusion in a harmonic potential:


\begin{equation}
C(t) = \frac{2 \sqrt{\pi } e^{-\frac{2 t}{\bar{\tau} }}}{\left(\frac{k}{k_B T}\right){}^{3/2} \sqrt{\frac{k_B T
   \left(1-e^{-\frac{4 t}{\bar{\tau} }}\right)}{k}} \sqrt{\frac{k \left(\coth \left(\frac{2 t}{\bar{\tau}
   }\right)+1\right)}{k_B T}}}
\label{aCt} 
\end{equation}
\noindent
Alternatively, we can apply equation \ref{geq}, and obtain a general closed form expression for approximating $C(t)$ by summing over a discrete configuration space of $N$ milestones rather than integrating over a continuous one:


\begin{multline}
C(t) = \frac{1}{\sqrt{\frac{2 \pi k_B T \left(1-e^{-\frac{4 t}{\bar{\tau}} }\right)}{k}}} \\ 
\sum _{i=1}^N x_i P_i(\infty)  \Delta  x
\sum _{j=1}^N \left( x_j Q_{ji}(t) \Delta x + x_i Q_{ii}(t) \Delta  x \right) \\
\label{gaCt} 
\end{multline}
\noindent
where 
\begin{multline}
Q_{ji}(t) = \exp \left( -\frac{k \left(\coth
   \left(\frac{2 t}{\bar{\tau} }\right)-1\right) \left(x_i-x_j e^{\frac{2 t}{\bar{\tau} }}\right)^2}{4 k_B T} \right) \\
Q_{ii}(t) = \exp \left( -\frac{x_i^2 k \tanh 
   \left(\frac{t}{\bar{\tau} }\right)}{2 k_B T} \right)
\label{trans} 
\end{multline}
\noindent
and $\Delta x$ is the distance between the evenly spaced milestones. $Q_{ji}(t)$ represents the discrete time-dependent probability density as a function of time that our system is in configuration $x_i$ at time $t$, given that the system was in state $x_j$ at time $t=0$. Likewise, $Q_{ii}(t)$ is the discrete probability density as a function of time that our system is still in configuration $x_i$ at time $t$ if it started in configuration $x_i$ at time $t=0$. Thinking in terms of the assumption of Markov statistics for transitions between milestones inherent to the Milestoning method, it makes sense that these probabilities are added given that we are interested in the outcome of finding our system in configuration $x_i$ whether it was already there, or it arrived there from another configuration. 

\begin{figure}[h]
\centerline{\epsfig{figure=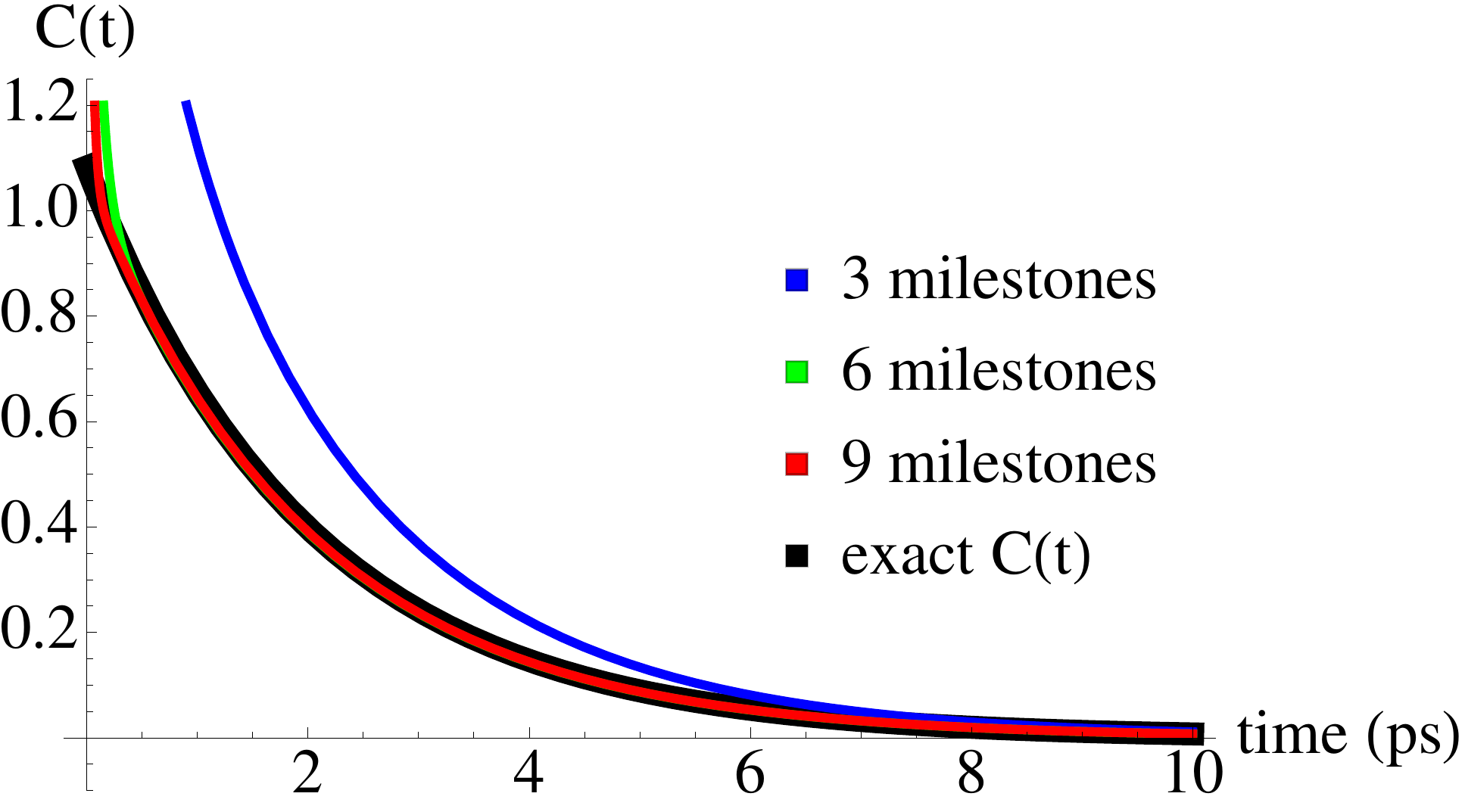, width=3.5in}}
\caption{This figure shows the approximate time correlation functions calculated using equation \ref{geq} for 3, 6, and 9 milestones overlaid on top of the exact analytical function $C(t)$.}
\label{hoc}
\end{figure}

The most straightforward and intuitive way to compare equations \ref{aCt} and \ref{gaCt} is to plot them. In figure \ref{hoc}, we can compare the exact time correlation function for diffusion in a harmonic potential (with parameters $\beta = .35, k = 5, \text{ and } D = .2857$) with the approximate $C(t)$ generated using equation \ref{gaCt}. Discretizing the space to three milestones is clearly too coarse of an approximation, but the gain in accuracy in going from 6 to 9 milestones is quite modest. As one might expect, the discrete approximation of the time correlation function is most accurate for long times and least accurate for $C(0)$. It turns out that this sacrifice in accuracy is a meager one because $C(0)$ is always available from Milestoning data because it is equivalent to the sum approximation of the variance in configuration space at equilibrium, $\sum_{i = 1}^N x_i^2 P_i(\infty)$. This will be leveraged to our advantage in the following section.     



\section{Numerical Demonstration}

\subsection*{1D Fokker-Planck Diffusion on a Bistable Potential}
\begin{figure}[h]
\centerline{\epsfig{figure=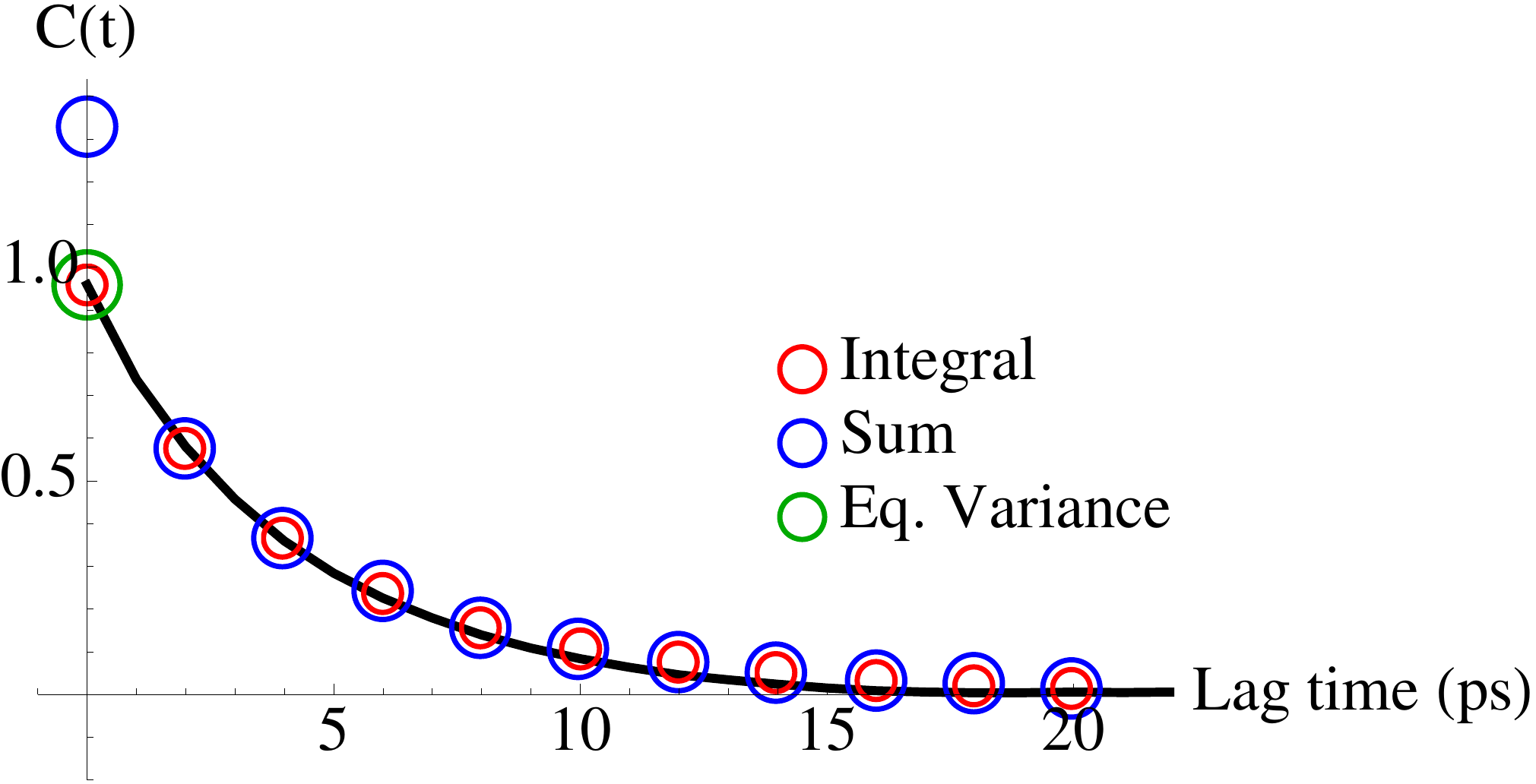, width=3.5in}}
\caption{This plot demonstrates a successful implementation of our method for approximating time correlation functions in continuous space by summing over time dependent joint probabilities of transitions between discrete states, as obtained in Milestoning simulations. The red rings mark the data points from implementing equation \ref{intr}, the blue data points indicate the positions where the full nested sum approximation of equation \ref{geq} was implemented, and the green ring is the data point for $C(0)$ calculated from equilibrium probabilities which is used to replace the value of $C(0)$ generated using equation \ref{geq}. The data is shown superimposed over the time correlation function $C(t)$, represented by a solid black line, calculated using the traditional method of equation \ref{trad}.}
\label{fokp}
\end{figure}

In order to further validate the approach of calculating time correlation functions using the nested sum in equation \ref{geq} in a discrete configuration space to approximate integrating equation \ref{coft} in continuous conformation space, the method was applied to a  simple two well potential of equation $y = (x-1)^2(x+1)^2$, where the time evolution of the probability density function in configuration space was calculated using a Fokker-Planck formalism:

\begin{equation}
\frac{\partial \rho(x,t)}{\partial t} = D \frac{\partial^2 \rho(x,t)}{\partial x^2} + \frac{D}{k_{B}T}\frac{\partial}{\partial x}(\rho(x,t)\frac{\partial V}{\partial x}) 
\label{fp}
\end{equation} 
\noindent
By repeatedly solving equation \ref{fp} numerically with the using the \textit{Mathematica} software package \cite{Mathematica}, using a normalized Gaussian distribution centered at the various $x_i(0)$ values as the initial condition, the manifolds $g(x, x_i(0), t)$ were obtained for each of the $10$ milestone configurations $x_i$ in the set $\{ -2, -1.6, ..., 1.6, 2 \}$. These manifolds were then used to find $C(t)$ using both the intermediate method described by equation \ref{intr} (shown as red circles in figure \ref{fokp}) as well as our fully developed discrete method described by equation \ref{geq} (shown as blue circles in figure \ref{fokp}). In the case of the equation \ref{intr}, the integral $\int x g(x,x(0),t) dx$ was numerically integrated directly, while in the case of equation \ref{geq}, the manifold $g(x,x(0),t)$ was used to obtain values of $P_i(x(0),t)$ by multiplying $g(x,x(0),t) \Delta x$, similar to the transformation from equation \ref{intr} to equation \ref{geq}, but in reverse. The results are shown superimposed over a plot of the time correlation function for the system obtained in the traditional manner by running $10^9$ steps of langevin dynamics and then calculating the time correlation function over this one long trajectory using the equation:

\begin{equation}
C(t) = \frac{1}{n - t} \sum_{i=1}^{n-t} x_i x_{t+i} 
\label{trad}
\end{equation} 
\noindent

We would like to point out that, as we alluded to in the previous section, the data point for $C(0)$ is the only portion of the time correlation function approximated using equation \ref{geq} with any appreciable error. In practice, the data point for $C(0)$ can always be replaced with the value obtained from the sum $C(0) = \sum_i x_i^2 P_i(\infty)$ (shown as the green ring in figure \ref{fokp}), due to the fact that the set of equilibrium probabilities, $P_i(\infty)$ are always known from Milestoning simulations. 

\subsection*{Random Walk / Path Integral Methodology}

\begin{figure}[h]
\centerline{\epsfig{figure=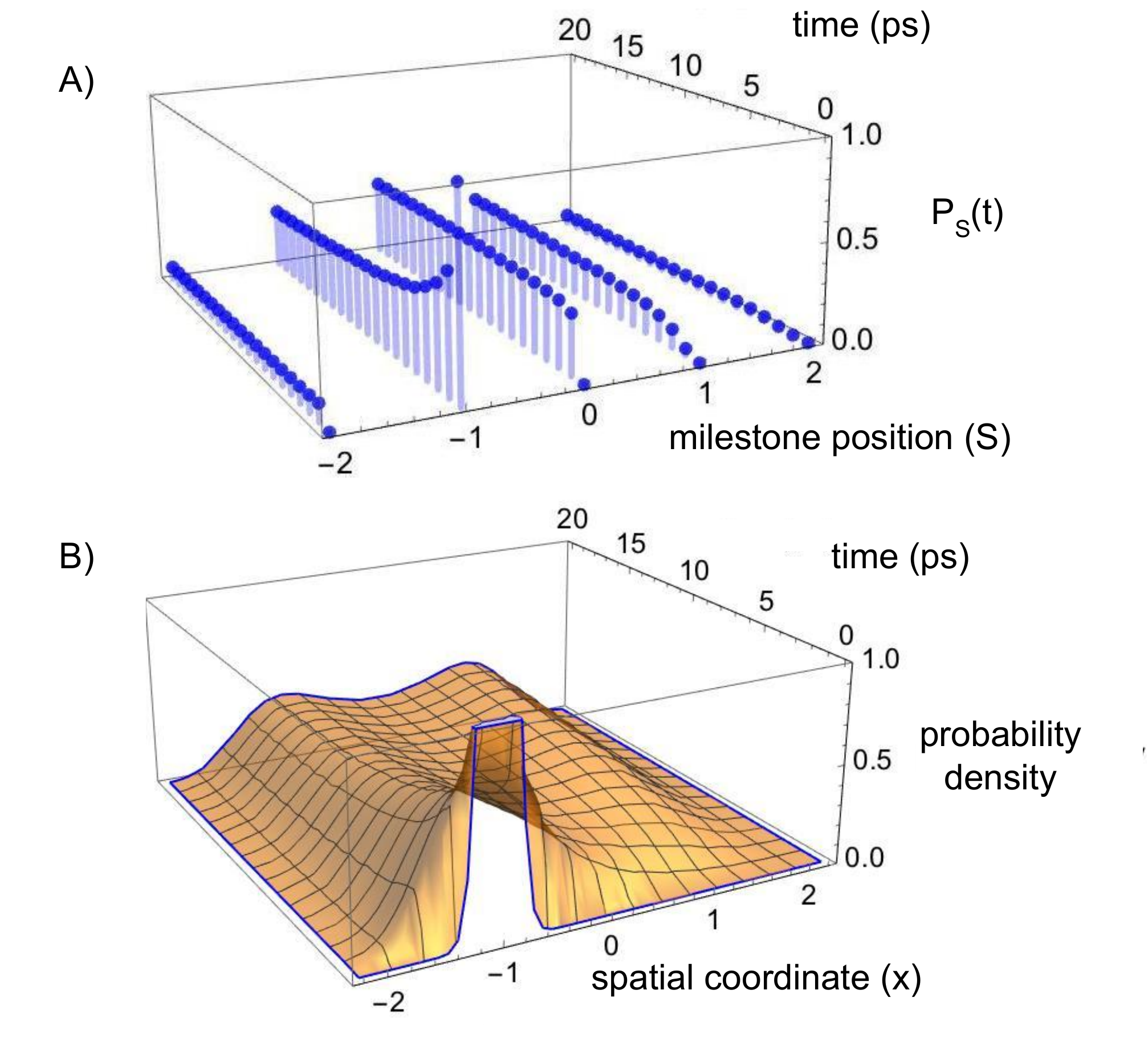, width=4.0in}}
\caption{This figure shows a graphical comparison between the time evolution of a discrete probability distribution for a set of 5 milestone configurations subjected to the two well 1D potential found in the Numerical Demonstration section using our random walk / path integral methodology (part A), and the manifold representing the time evolution of a continuous probability density function of configurations for the same two well system subjected to Fokker-Planck diffusion (part B).  Part A is the set of probabilities as a function of time for the system being found at each milestone configuration, given that the system was in configuration $x = -1$ at time $t = 0$, and part B shows Fokker-Planck diffusion on the same two well system. Note that the random walk in part A began at the milestone located at $x = -1$2, thus we see a decay from $\{ P_1(0) = 0, P_2(0) = 1, P_3(0) = 0, P_4(0) = 0 , P_5(0)\}$ to the equilibrium distribution, the same way our initial continuous distribution, a normalized Gaussian centered at $-1$, decays to the equilibrium probability distribution predicted by the Bolzmann distribution for the two well potential, and both evolve in time on about the same time scale.}
\label{surfs}
\end{figure}

\begin{figure}[h]
\centerline{\epsfig{figure=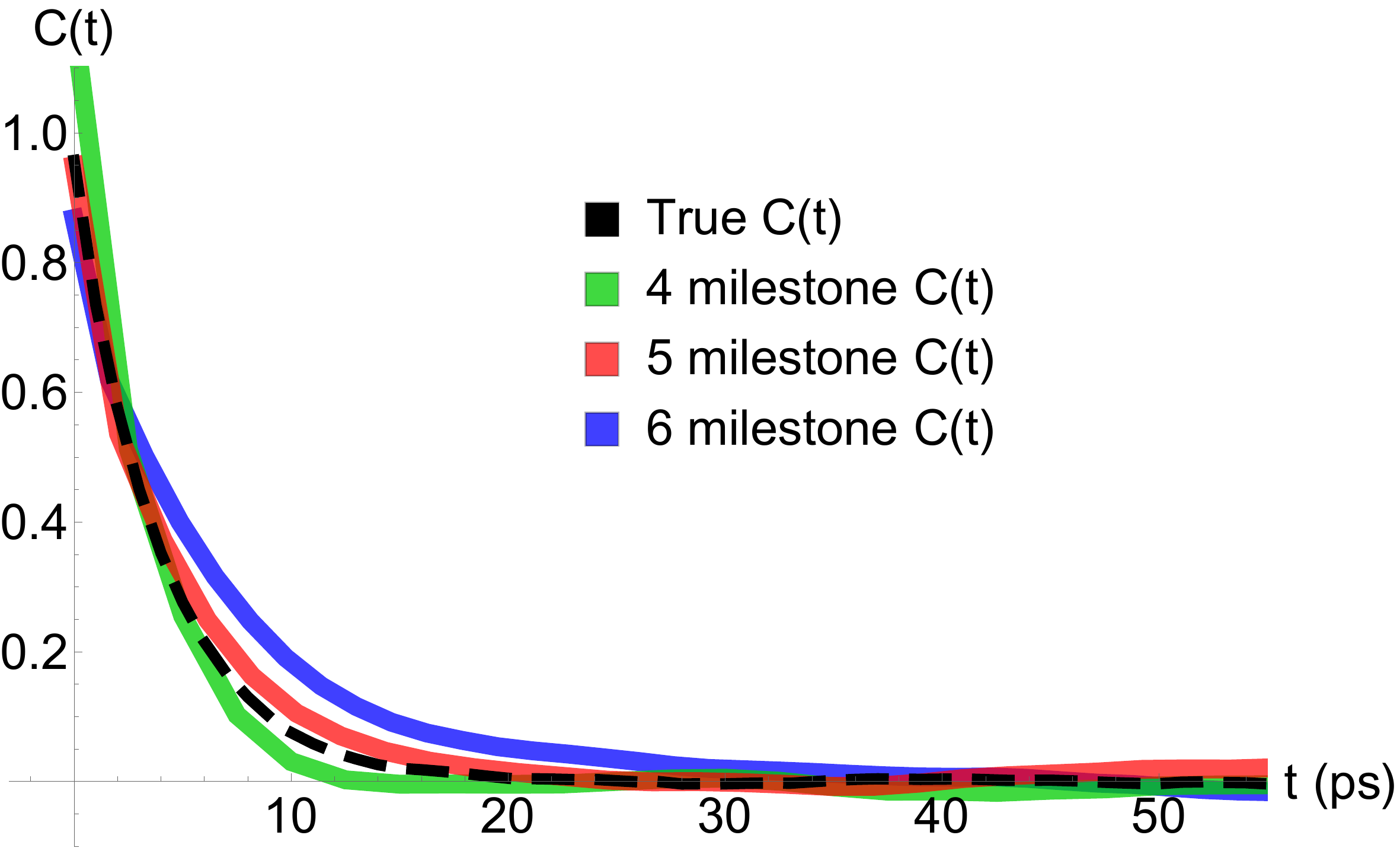, width=3.5in}}
\caption{Shown here are time correlation functions which were calculated by first generating one long random walk using the method introduced in this article, then linking each point in the trajectory using linear interpolation, and finally using equation \ref{trad} to calculate $C(t)$.}
\label{ctTerp}
\end{figure}

In order to make use of the formalism for obtaining autocorrelation in a discrete configuration space, as introduced in the Theory section, we require an expression for $P_s(t | x_i(0))$, i.e. the probability that our system is in configuration $s$ at time $t$, given that it was in configuration $i$ at time $t = 0$. Since previous implementations of the milestoning method have been ``based on iterative determination of stationary flux vectors at milestones" \cite{exactMstone}, and not the determination of non-equilibrium time dependent fluxes given some initial configuration, it was necessary to devise a methodology for obtaining the function $P_s(t | x_r(0))$ from milestoning data. In the case of diffusive systems which can be described using a Fokker-Planck formalism (eq. \ref{fp}), the Fokker-Planck equation can be solved for a manifold $\rho(x,t)$ which represents a probability density of configurations evolving in time, where the distribution at time $t = 0$ is the distribution dictated by the initial condition and the distribution as $t \rightarrow \infty$ is equivalent to the equilibrium distribution in x. While this Fokker-Planck description can be directly solved for the time evolution of a probability density function of configurations (when tractable, as in figure \ref{fokp}), it is also possible to obtain the manifold $\rho(x,t)$ via a path integral approach using a large ensemble of trajectories generated using stochastic models such as Langevin dynamics. This equivalence was the inspiration behind the random walk / path integral method introduced in this section. There are some differences however, for example, instead of Langevin trajectories, we use random walks along the given set of milestones. Very long random walks, orders of magnitude longer than time scales accessible to molecular dynamics, can be quickly generated with minimal computational cost by taking advantage of two data sets which are already known in any milestoning calculation: the transition matrix \textbf{K} (essentially a Markov matrix) and the set of all $K_{AB}(\tau)$ functions, which are the probability density functions of transition times between milestone $A$ and milestone $B$. The $K_{AB}(\tau)$ functions are obtained by histogramming transition times between milestones, and each element $K_{ij}$ of the matrix \textbf{K} is obtained by integrating the distributions of transition times, $k_{ij}(\tau)$, over all time $\tau$ and then normalizing each row to impose the constraint that the system at state $i$ has probability 1 of transitioning to one of the states to which it is coupled ($j$). Since the matrix \textbf{K} gives the equilibrium transition probabilities between milestones, and the $k_{ij}$ functions are probability density functions for the transition time between connected milestones, these two pieces of information can be used to construct time-dependent random walks along a set of milestones. Each step taken from some current configuration $i$ is chosen by selecting between each possible coupled state $j$, weighted by the transition probabilities from \textbf{K}, next, the amount of time each selected transition from state $i$ to $j$ took is selected randomly from the distribution defined by $k_{ij}(\tau)$. In this manner, trajectories of arbitrary length in this discrete space can be very quickly generated in only the amount of CPU time necessary to select $2N$ random numbers, where $N$ is the desired number of steps in the random walk. Once a large set of these random walks is generated, they can be used to calculate discrete versions of the same $\rho(x,t)$ manifolds which would be obtained as the solutions to the Fokker-Planck equation (see figure \ref{surfs}). To elaborate on this, consider a single random walk along the milestone configurations. If, at each time step, we histogram the frequency with which our system has visited each milestone configuration up to that point in time into a normalized distribution, then we have constructed a discrete manifold in configuration space $x$ and time $t$ which represents the time evolution of the probability distribution of finding our system in a particular configuration for this particular realization of a random walk in our discrete configuration space. From here, it only remains to average the set of probability distributions generated from numerous manifestations of the random walk. An alternative approach to calculating time correlation functions from these random walks would be to ``connect the dots'' along the random walk using an interpolation method, and then use the traditional approach to numerically calculating time correlation, shown in equation \ref{trad}, from the resulting continuous function, as shown in figure \ref{ctTerp}.   

\begin{figure}[h]
\centerline{\epsfig{figure=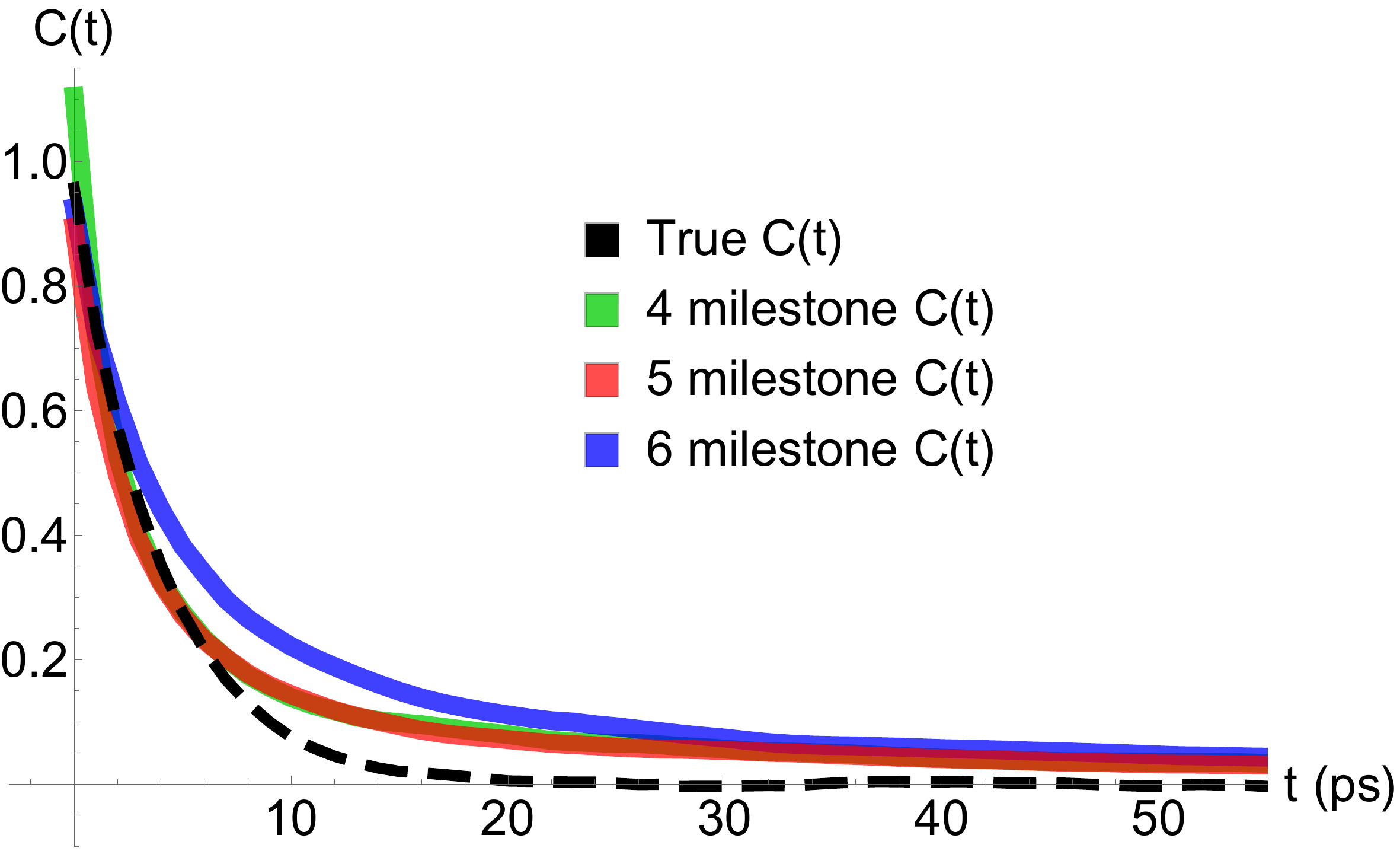, width=3.5in}}
\caption{Shown here are time correlation functions calculated using equation \ref{geq}, where the conditional probability as a functions of time, $P_s(t | x(0))$, are calculated using our random walk / path integral methodology, represented graphically in figure \ref{surfs}A.}
\label{ctdist}
\end{figure}

\section{Application to Calculating Long-Time RDCs in Atomistic Simulations}

\subsection*{Application of Discrete Space Time Correlation Methodology to the Alanine Dipeptide Bond Vector}

\begin{figure}[h]
\centerline{\epsfig{figure=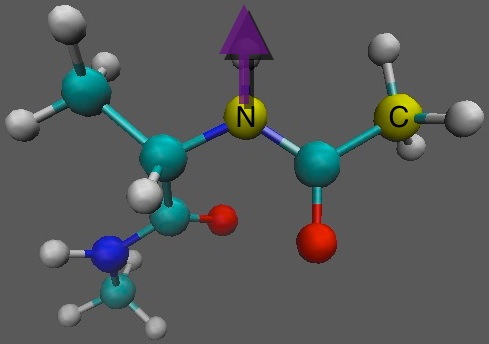, width=2.9in}}
\caption{Shown in this figure is the alanine dipeptide molecule used as our model system. The two atoms shown in yellow were held fixed in space while the rest of the molecule was subjected to Langevin dynamics. The purple arrow gives the orientation of the bond vector which served as the measurable in our time correlation function calculations. }
\label{mol}
\end{figure}

\begin{figure}[h]
\centerline{\epsfig{figure=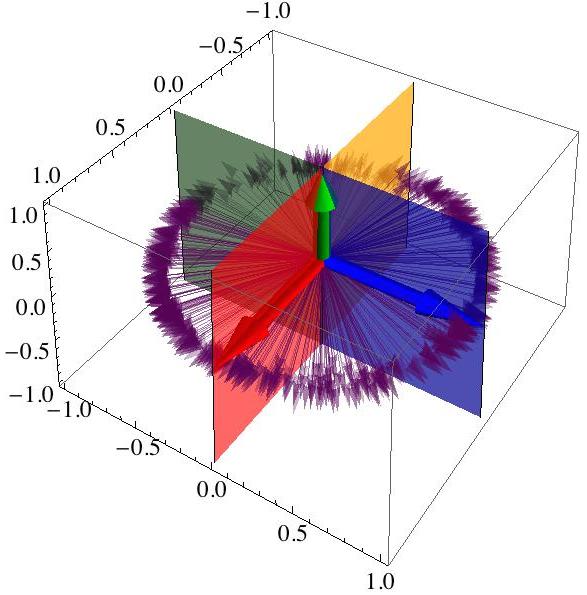, width=3.5in}}
\caption{Shown here is a graphical representation of the four milestone configuration for measuring the time correlation function of the alanine dipeptide bond vector. Although the bond vector, shown as many thin, purple arrows, posses three degrees of freedom as it fluctuates in time, we are able to choose a frame of reference where the bulk of the motion is taking place as a rotation about the z-axis, shown as a thick green arrow. Using the four milestones, shown as the red, green, yellow, and blue planes, we can calculate transition time probability distributions between each pair of adjacent milestones.}
\label{planes}
\end{figure}

In this section, we describe an application of our methodology to a molecular system. Shown in figure \ref{mol} is the molecular structure of our system, alanine dipeptide. After constraining the nitrogen and carbon atoms labeled in yellow to remain fixed at their initial positions, Langevin dynamics at $T = 300 K$ was run for $4 \times 10^7$ time steps with a time step size of 0.001 ps for a total of 40 nanoseconds using the CHARMM molecular dynamics software package. As the molecular dynamics simulation ran, the orientation of the bond vector extending from the center of the labeled nitrogen atom to the center of the hydrogen atom indicated by the purple arrow in figure \ref{mol} was recorded. Although this bond vector possesses three spatial degrees of freedom, it's orientation could be well approximated by a single rotational degree of freedom, as shown in figure \ref{planes}. By counting the number of time steps between transitions from one milestone state to the next (shown graphically as the four colored planes in figure \ref{planes}) over the course of the 40 nanosecond trajectory, probability distribution functions for the transition times between neighboring pairs were constructed as histograms to obtain the set of $k_{ij}(\tau)$ functions for each pair of neighboring milestone states. These $k_{ij}(\tau)$ functions were then used as the basis for the random walk / path integral approach described in the previous section. Thusly, the $P_s(t | x_0)$ functions necessary to calculate the time correlation function using equation \ref{L2Coft} were calculated by averaging 75,000 different time-dependent probability distribution functions which each resulted from some particular manifestation of the random walk. The time correlation functions of interest for this system are those which can be calculated using the Lipari-Szabo formalism \cite{lipari}, as implemented by Xing and Andricioaei \cite{Xing}, using the equation:

\begin{equation}
C(t) = \langle L_2(\textbf{u}(0)\textbf{u}(t)) \rangle 
\label{L2}
\end{equation} 

\noindent
where $L_2(\textbf{u}(0)\textbf{u}(t))$ refers to plugging the scalar resulting from the dot product of time series measurements of the bond vector \textbf{u} into the second order Legendre polynomial. This motif of measuring the autocorrelation of this value is then applied to equation \ref{geq} to yield the discrete space time correlation function relationship:

\begin{equation}
C(t) = \sum_i L_2\left[\sum_s (\textbf{u}_i(0) \cdot \textbf{u}_s) P_s(t | \textbf{u}_i(0)) \right] P_i(\infty)
\label{L2Coft}
\end{equation} 
\noindent
where the vectors $\textbf{u}_i$ represent the different possible values for the bond vector, given the coarse graining of the bond vector into a discrete space. The oscillatory and slower decay in correlation for the 4 milestone case is an effect of coarse graining the space. This is due to a loss in entropy in going from the continuous space to the discrete one, i.e. if only four possibilities exist for the position of the bond vector, the probability of pointing in the same direction as that of a previous time step increases compared to a system where 8 or more configurations are possible.

Notably, the oscillatory and slower decay in correlation for the 4 milestone case is an effect of coarse graining the space (the oscillations are reproduceable). This is due to a loss in entropy in going from the continuous space to the discrete one, i.e. if only four possibilities exist for the position of the bond vector, the probability of pointing in the same direction as that of a previous time step increases compared to a system where 8 or more configurations are possible. 


\begin{figure}[h]
\centerline{\epsfig{figure=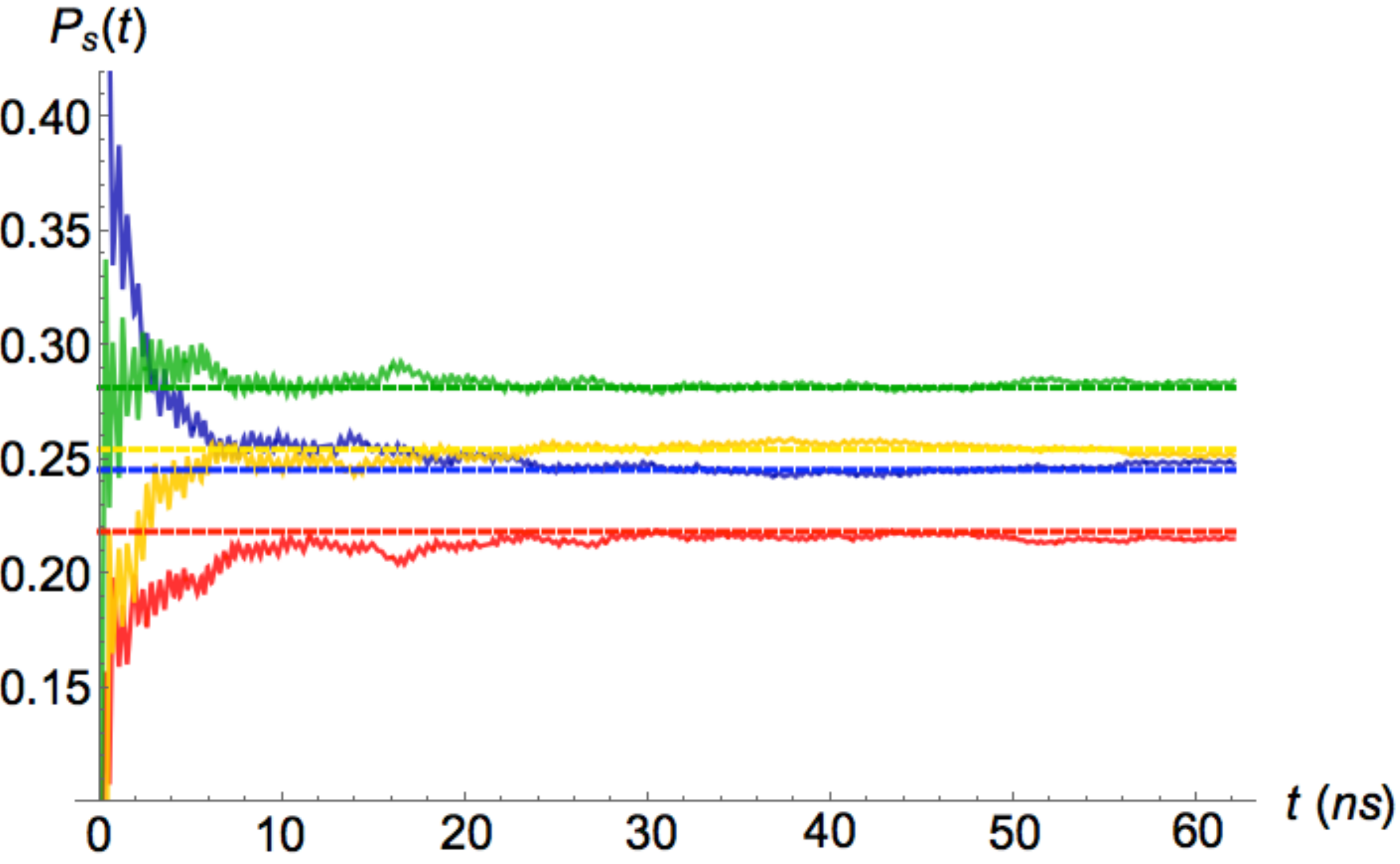, width=3.5in}}
\caption{This plot gives the probability of finding our system in each of the four milestone configurations as a function time, given that we began the simulation with our system in the configuration shown as the blue plane, using the same color scheme as in figure \ref{planes}. The probability of being found in the blue milestone is equal to 1 at time $t = 0$ of course, but the plot range stops shy of $P_s(t) = 1$ in order to provide a more detailed view. Note that the probability of the system being in any of the other three milestone configurations is equal to zero at time $t = 0$, as expected. These functions were calculated using the methodology described in the Random Walk / Path Integral Methodology section. These functions contributed to the calculation of $C(t)$ shown in figure \ref{alaCoft}. Note that the probabilities converge to their equilibrium values on roughly the same timescale that $C(t)$ converges to its long time value.}
\label{psPlot}
\end{figure}

\begin{figure}[h]
\centerline{\epsfig{figure=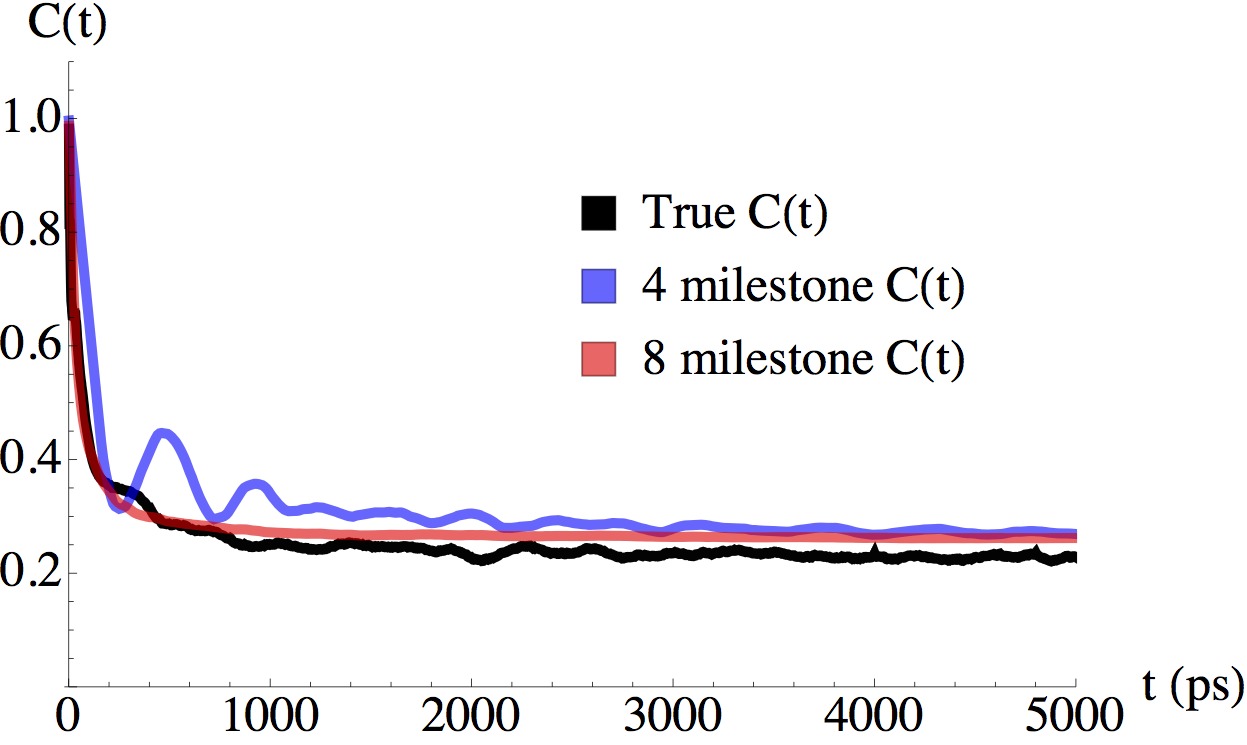, width=3.7in}} 
\caption{This figure shows the approximate time correlation functions calculated using equation \ref{geq} superimposed over the true time correlation function, calculated using equation \ref{trad}. The 4 milestone $C(t)$ function was calculated with the milestones placed 90 degrees apart as illustrated in figure \ref{planes}, while the 8 milestone configuration was the same motif, only with 8 planes placed 45 degrees apart.}
\label{alaCoft}
\end{figure}

\section{Concluding Discussion}
We have demonstrated for the first time that time correlation functions for continuous processes can be approximated using equation \ref{geq} to coarse grain the configuration space to a discrete one. Additionally, we have introduced a novel method for extending milestoning into non-equilibrium regimes by numerically calculating the time-dependent fluxes $P_s(t|x_i(0))$. The method consists of constructing random walks in the discrete configuration space, defined by a set of milestone configurations, from transition time probability density functions $k_{ij}(\tau)$ obtained using the milestoning method, followed by calculating time-dependent histograms of milestone states occupied using the stochastic path integral method described in the Random Walk / Path Integral Methodology section.

The time correlation function for the harmonic oscillator calculated analytically using our discretization method showed excellent agreement with the true time correlation function $C(t)$, also obtained analytically, for a harmonic oscillator. There was also an excellent agreement between the $C(t)$ calculated for a discrete configuration space for a bistable potential and the true autocorrelation function, where  $P_s(t|x_i(0))$ was obtained by numerically solving a Fokker-Planck equation. We also obtained a promising result from applying the discretization method of equation \ref{L2Coft} in conjunction with the stochastic path integral method to an atomistic system. The autocorrelation function $C(t)$ for the bond vector calculated using the methods introduced herein showed a nice agreement with the true $C(t)$ calculated using equation \ref{L2}. The limitations to the methods we have introduced appear to be limited to the challenges inherent to implementation of the milestoning method. A key advantage of our method is that the random walks between discrete configurations can be constructed at trivial computational cost, allowing for us to make predictions well into time regimes inaccessible to molecular dynamics simulations. We would like to note that, although the calculations described in this article were performed on systems where the observable of interest was constant along each milestone hyperplane, the method can easily be generalized for systems where the observable varies along each milestone hyperplane. In order to account for such observables, one must simply construct equilibrium probability distributions of the observable on each hyperplane, then select from this distribution at each time step of the random walk along the milestones. In other words, at each step, the algorithm must first choose the next step to take using the transition matrix, then select the transition time from the appropriate transition time distribution function, then select the value of the observable from the probability distribution describing the observable along that hyperplane. We feel that the methods introduced in this paper have the potential to allow for the calculation of experimental observables from molecular dynamics simulations that are currently unattainable by brute force long time simulations. The method presented herein could also be further enhanced by combining it with the enhanced sampling methodology introduced in the companion article to this paper, also found within this publication \cite{grazioli1}. 
\section{Acknowledgments}  IA acknowledges funds from an NSF CAREER award (CHE-0548047).


\clearpage

\bibliographystyle{unsrt}
\bibliography{tcms}

\end{document}